\documentclass[conference]{IEEEtran}
\usepackage{cite}
\usepackage{amsmath,amssymb,amsfonts}
\usepackage{algorithmic}
\usepackage{graphicx}
\usepackage{textcomp}
\usepackage{xcolor}
\usepackage{subcaption}
\usepackage{pgfplots}
\usepackage{verbatim}
\usepackage{flushend}
\def\BibTeX{{\rm B\kern-.05em{\sc i\kern-.025em b}\kern-.08em
    T\kern-.1667em\lower.7ex\hbox{E}\kern-.125emX}}

\begin{document}

\textheight=24.5cm

\title{From Self-Adaptation to Self-Evolution \\Leveraging the Operational Design Domain}


\author{\IEEEauthorblockN{Danny Weyns}
\IEEEauthorblockA{\textit{Department of Computer Science} \\
KU Leuven, Belgium and   
Linnaeus University, Sweden \\
danny.weyns@kuleuven.be\vspace{-10pt}}
\and
\IEEEauthorblockN{Jesper Andersson}
\IEEEauthorblockA{
\textit{Department of Computer Science and Media Technology}\\
Linnaeus University, Sweden \\
jesper.andersson@lnu.se\vspace{-10pt}}
}

\maketitle


\begin{abstract}
Engineering long-running computing systems that achieve their goals under ever-changing conditions pose significant challenges. Self-adaptation has shown to be a viable approach to dealing with changing conditions. Yet, the capabilities of a self-adaptive system are constrained by its operational design domain (ODD), i.e., the conditions for which the system was built (requirements, constraints, and context). Changes, such as adding new goals or dealing with new contexts, require system evolution. While the system evolution process has been automated substantially, it remains human-driven. Given the growing complexity of computing systems, human-driven evolution will eventually become unmanageable. In this paper, we provide a definition for ODD and apply it to a self-adaptive system. Next, we explain why conditions not covered by the ODD require system evolution. Then, we outline a new approach for self-evolution that leverages the concept of ODD, enabling a system to evolve autonomously to deal with conditions not anticipated by its initial ODD. We conclude with open challenges to realise self-evolution. 
\end{abstract}


In essence, self-adaptation is an approach that enables a system to deal with uncertainty and business continuity autonomously, or with minimal human support if needed. The key idea of self-adaptation is to let the system gather new knowledge at runtime to resolve uncertainties, reason about itself, its context, and goals, and adapt to realise a set of adaptation goals. Since the pioneering work of Oreizy et al.~\cite{oreizy1999aba}, Kephart and Chess~\cite{kephart2003vision}, and Garlan et al.~\cite{garlan2004rainbow} 
the field has matured and generated a large body of knowledge~\cite{weyns2021introduction}. A recent large-scale survey~\cite{3524844.3528077,survey-report} provided evidence that self-adaptation is also widely applied in industry. 

Although self-adaptation has shown to be a viable approach to dealing with changes, the conditions self-adaptive systems can handle are determined by their design. In this paper, we use the concept of \text{operational design domain} -- ODD~\cite{SAE} to determine the conditions a computing system can handle by its design, hence also a self-adaptive system. These conditions comprise the requirements the system should realise, the constraints that need to be taken into account, and the context in which the system should operate. 
Any conditions outside the ODD, meaning requirements, context, and constraints that are not considered during the initial system design and hence are not covered by the OOD, require system evolution~\cite{Chapin2001,Buckley2005}. 

While the process of system evolution has been automated substantially in the past decade, it remains, in essence, human-driven. Given the growing complexity of computing systems, human-driven evolution will eventually become unmanageable~\cite{Reussner2019,Andersson2013}. A key challenge in realising systems that can evolve themselves is to determine and specify the \textit{evolution target}. The evolution target 
enables a system to assess and select
decision alternatives to evolve.
In particular, the evolution target provides a stopping criterion determining when the system meets the new or changing requirements, constraints, and contexts. In this paper, we propose the ODD as a representation of an evolution target that enables a system to self-evolve to deal with conditions initially not anticipated. 

The remainder of this paper is structured as follows. We start with motivating the need for self-evolution and highlighting other relevant work. Then we introduce the concept of ODD and apply it to a self-adaptive system. Next, we explain why conditions not covered by the ODD require system evolution. Then, we outline a new architecture for self-evolving systems leveraging the ODD as an evolution target. Finally, we look at open challenges to realise self-evolving systems.

\section{Motivation for Self-Evolution \& Related Work}

The software evolution process has been developing towards full automation during the past decade~\cite{978-3-319-08738,6802994}, yet, it remains, in essence, human-driven. With the increasing complexity of computing systems, the rapid advancements of technologies, the need for systems operating 24/7, and the ever-changing demands on computing systems, human-driven evolution will eventually become unmanageable~\cite{10.1145-336512.336534,10.1109-FOSE.2007.20,Reussner2019}. Hence, there is a need for systems that can evolve themselves. 

We highlight a selection of fields that have contributed knowledge for tackling the problem of self-evolution. A classic field is autonomous systems~\cite{Tzafestas2012,Paulovich2018,Wooldrige2009} that enable systems to operate independently of direct human supervision. Smart systems~\cite{6857843,7433937,8010538,9599303,JID210010} are equipped with advanced technologies to observe, reason, and act in intelligent ways, such as in Industry 4.0, smart grids, and transportation systems. 
The field of self-adaptation developed a body of knowledge on handling uncertainties at runtime~\cite{Esfahani2013,MAHDAVIHEZAVEHI201745,10.1145/2797433.2797497,9196226}. Mitigating such uncertainties calls for approaches that blend system engineering with system operation~\cite{1882362.1882367,abs-1903-04771,3524844.3528052}. Other related lines of work include self-improving system integration~\cite{BELLMAN2021136}, self-development~\cite{9909454}, and anomaly and novelty detection~\cite{GRUHL2021138}. While all these approaches offer stepping-stones towards self-evolving systems, an integrated perspective on self-evolution remains an open research challenge. 

\section{ODD of a Self-Adaptive System}

This section outlines self-adaptation and operational design domain (ODD). Further, we explain the ODD of a self-adaptive system and the ODD for a system undergoing evolution. 

\subsection{Self-Adaptation}

We refer to a self-adaptive system as one that complies with two fundamental principles of self-adaptation that are widely accepted and described in~\cite{weyns2021introduction}. The external principle states that a self-adaptive system can achieve its goals in the face of changes and uncertainties without or with minimal human intervention. While a human operator manages the uncertain operating conditions in a regular computing system, a self-adaptive system can deal with uncertainties autonomously, taking a set of adaptation goals as input. 
%
%
%
The internal principle states that a self-adaptive system comprises two distinct parts: the first part (managed system) interacts with the environment and is responsible for the domain concerns of the users for which the system is built; the second part (managing system) consists of a feedback loop that interacts with the first part (and monitors its environment) and is responsible for the adaptation concerns, i.e., concerns about the domain concerns. A human operator may be involved in realising the tasks of the managing system (human-in-the-loop) or observe the system in operation and only take action when needed (human-on-the-loop). 
%
%
%
\vspace{5pt}\\\textbf{Example.}
Let us consider an IoT system based on the DeltaIoT artifact~\cite{Iftikhar2017} as an example. The managed system consists of a set of battery-powered motes that periodically measure parameters in the environment (temperature, presence of people, etc.). The motes use multi-hop wireless communication to transmit the collected data to a central gateway. For simplicity, we consider the variable interference of the network (due to weather conditions, etc.) as the only uncertainty. 
The adaptation goals are to deliver packets with a packet loss below a given threshold while maximising motes battery lifetime. To achieve the throughput of packets under variable network interference, the gateway is extended with a managing system that can adapt the power setting of motes in three levels: minimum, medium, and maximum. A high power setting will minimise packet loss, yet, the energy consumption will be high, draining the batteries, and vice versa.   


\subsection{Operational Design Domain}

A computing system's operational design domain (ODD) is a loosely defined concept in the literature~\cite{9304552}. In the field of autonomous vehicles, the ODD is often used for safety assurance~\cite{Czarnecki}. There, the ODD typically refers to the specific domain in which an autonomous vehicle is designed to operate, including roadway types, speed ranges, environmental conditions, and domain constraints. In particular, the ODD is used to ensure that a vehicle will operate safely within defined conditions, detect when the conditions no longer hold and reach a safe state if that occurs~\cite{Hawkins,9304552,DBLP:conf/aaai/KoopmanF19}. 

We define the ODD of a system as the conditions for which the system was built, i.e., (i) the requirements $R$ the system should achieve (i.e., functional and quality requirements), (ii) the constraints 
$B$  that need to be taken into account when the system is built, and (iii) the context $C$ in which the system should operate. More precisely, we define the $ODD_{S}$ of a system $S$ as the set of working points $(u,c)$ with $u \in U$ the utility provided by the system in an operational context $c \in C$ that satisfies the design conditions $(R,B,C)$: 
%
%
%
\vspace{5pt}\\
 \mbox{\ \ }$ODD_{S} = \{(u,c)~|~(u,c).sat(R,B,C)_{S}\,\land\,c \in C\,, u \in U\}$
\vspace{5pt}\\
We use utility $u$ as an abstraction that expresses the usefulness of the system in satisfying the requirements.  

\begin{figure}
    \centering
    \includegraphics[width = 0.9\linewidth]{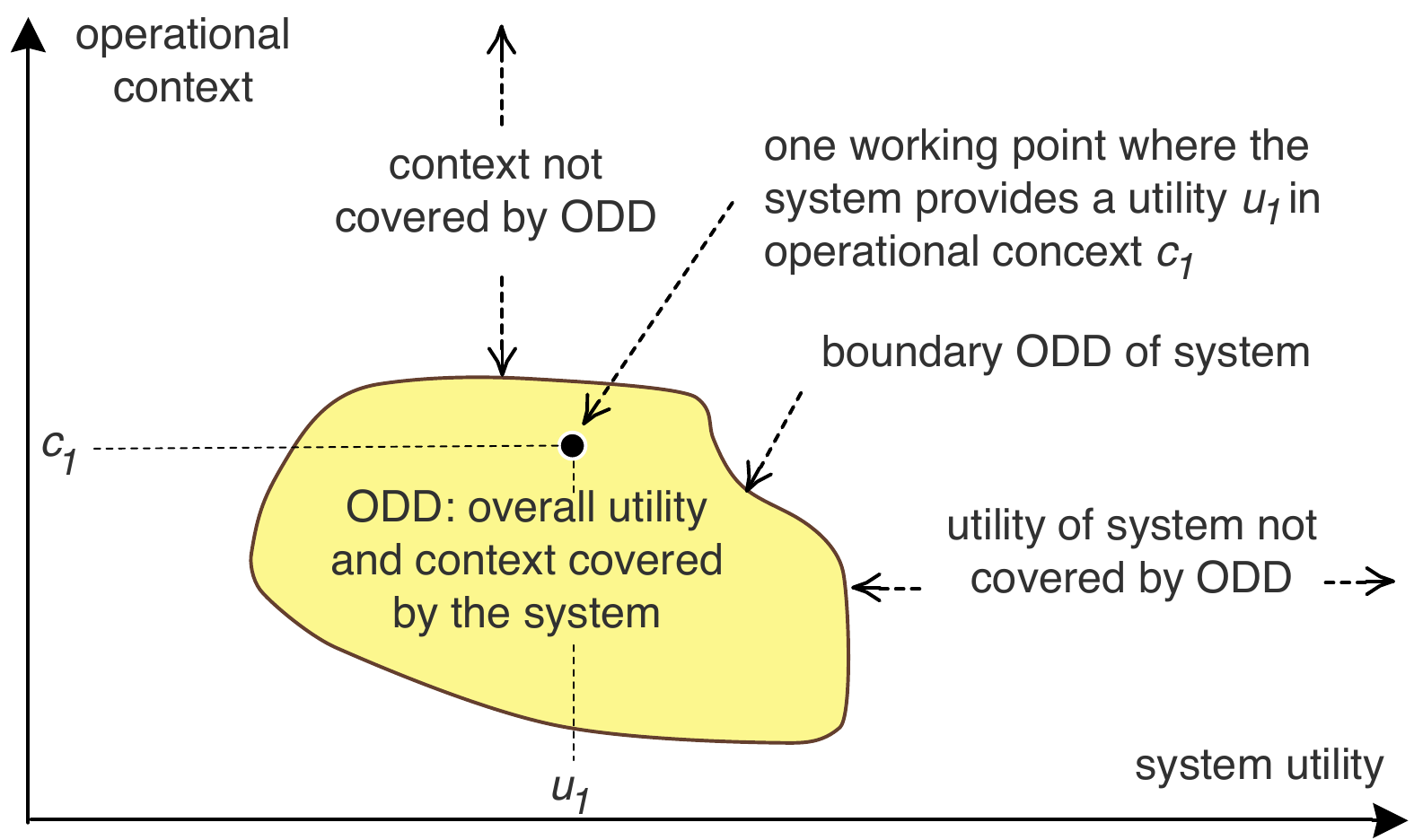}
    \caption{Illustration of ODD of a computing system}
    \label{fig:ODD}\vspace{-10pt}
\end{figure}

Figure~\ref{fig:ODD} provides a simple representation of an ODD. The shaded area covers the overall utility and operational context covered by the system. Hence, any working point of the system that is located within the boundary of the ODD of the system (i.e., within the shaded area) represents a context in which the system can provide a utility that satisfies the requirements and constraints. The system does not cover any point outside the boundary of the ODD (i.e., outside the shaded area). Note that in this simple example, the utility and context are expressed as simple scalars. In most real systems, the utility and context will be multi-dimensional. Yet, for the sake of simplicity, we use a two-dimensional representation without losing generality.  

The ODD may be defined explicitly by its stakeholders. However, it is often difficult or impossible to define the ODD completely for a real-world system. For instance, testing will guarantee the utility and system context for the tested parts of the ODD. 
Another example is a system that includes learning elements, for which it may be challenging to determine the ODD upfront precisely. For instance, a neural network can predict structures outside the training set once trained on the data. It may not be feasible to guarantee the precision of these predictions completely. Nevertheless, any piece of software, including machine learning-based solutions, has an ODD based on its design. Hence, the ODD is an inherent characteristic of a computing system, regardless of whether it is explicitly defined or not. Similarly, any computing system has a software architecture, explicitly defined or not~\cite{Bass2012}.

In essence, any computing system comprises a computation and
a domain model~\cite{maes1987concepts,FORMSj}. The computation represents the application logic of the system that realises the requirements. The domain model represents the context in which the system operates. In our setting, computing over the domain model reflects how a system ''navigates'' the ODD depending on the utility demands and the present context. 
Nevertheless, in most practical systems, a system may offer the option to select between different configurations. It is then the task of an operator to switch between different configurations while the system is in operation. The need for such switches depends on the conditions at hand. The ODD for each configuration covers a particular part of the overall ODD of the system. 

\begin{figure}
    \centering
    \includegraphics[width = \linewidth]{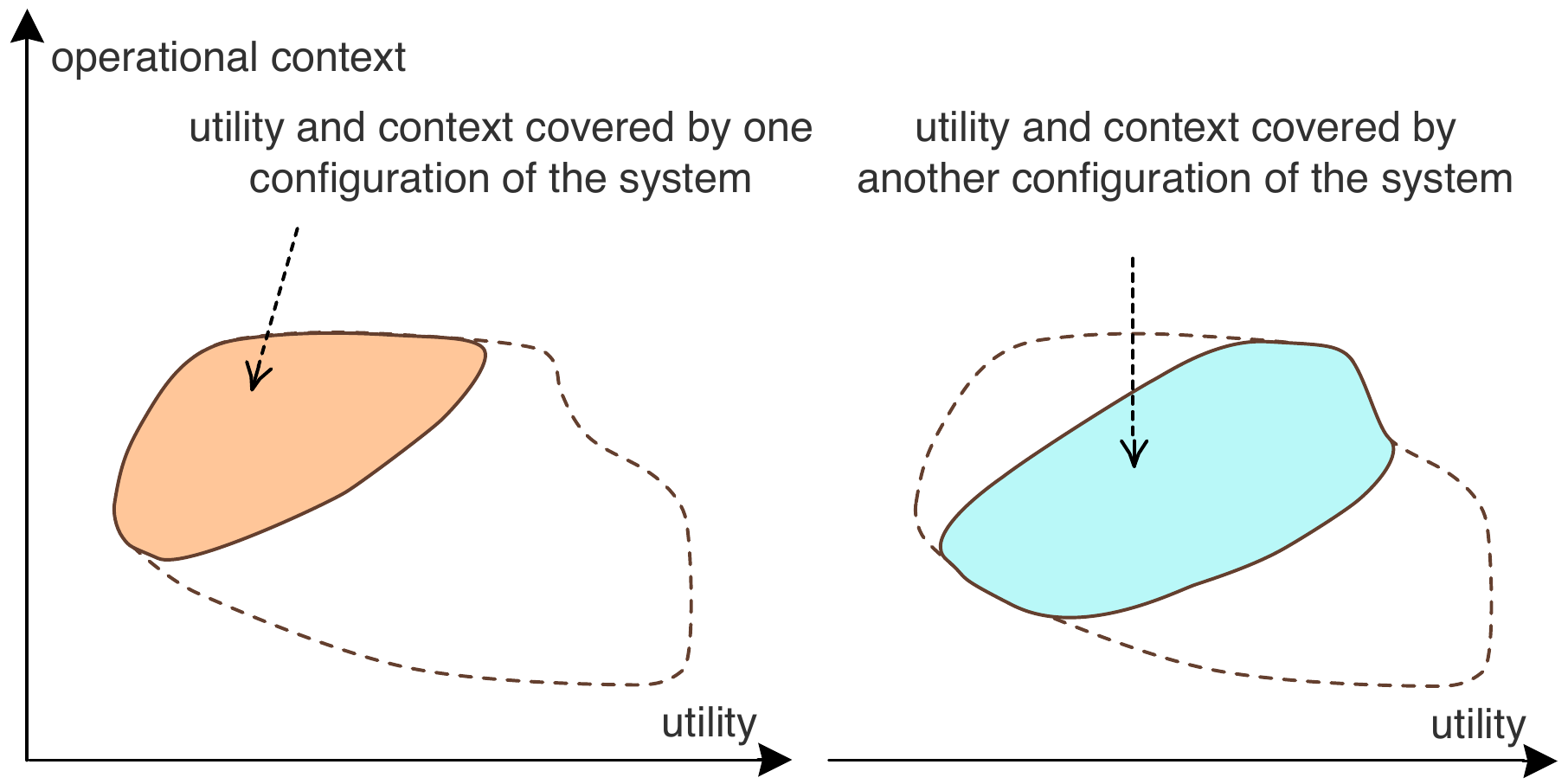}
    \caption{Illustration of ODD for two system configurations}
    \label{fig:ODD-1-2-separated}
\end{figure}

\begin{figure}
    \centering
    \includegraphics[width = 0.95\linewidth]{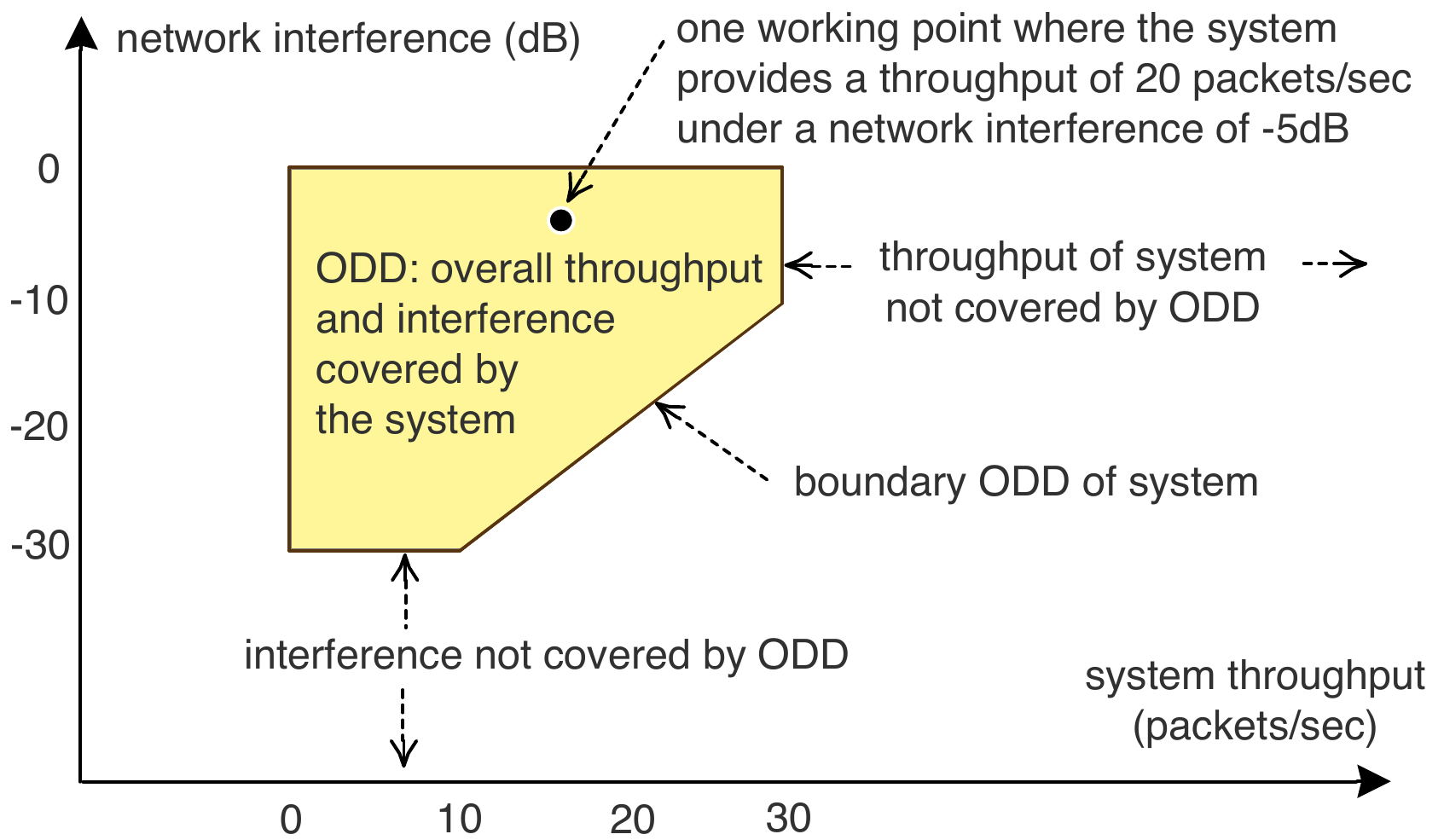}
    \caption{ODD of a simple IoT system}
    \label{fig:ODD-IoT}\vspace{-10pt}
\end{figure}

Figure~\ref{fig:ODD-1-2-separated} illustrates the parts of the ODD covered by two system configurations. The configuration parts may partially overlap. A system's overall utility and context is the union of the ODD parts covered by the system configurations.  
\vspace{5pt}\\\textbf{Example.}
Let us now consider the ODD of the simple IoT system we introduced above, 
shown in Figure~\ref{fig:ODD-IoT}. The system design supports a throughput of 10 packets per second for up to -30dB network interference. The throughput gradually decreases by lower network interference. The maximum throughput of 30 packets per second is achievable for a network interference between 0 and -10dB. 

\begin{figure*}
    \centering
    \includegraphics[width = 0.9\textwidth]{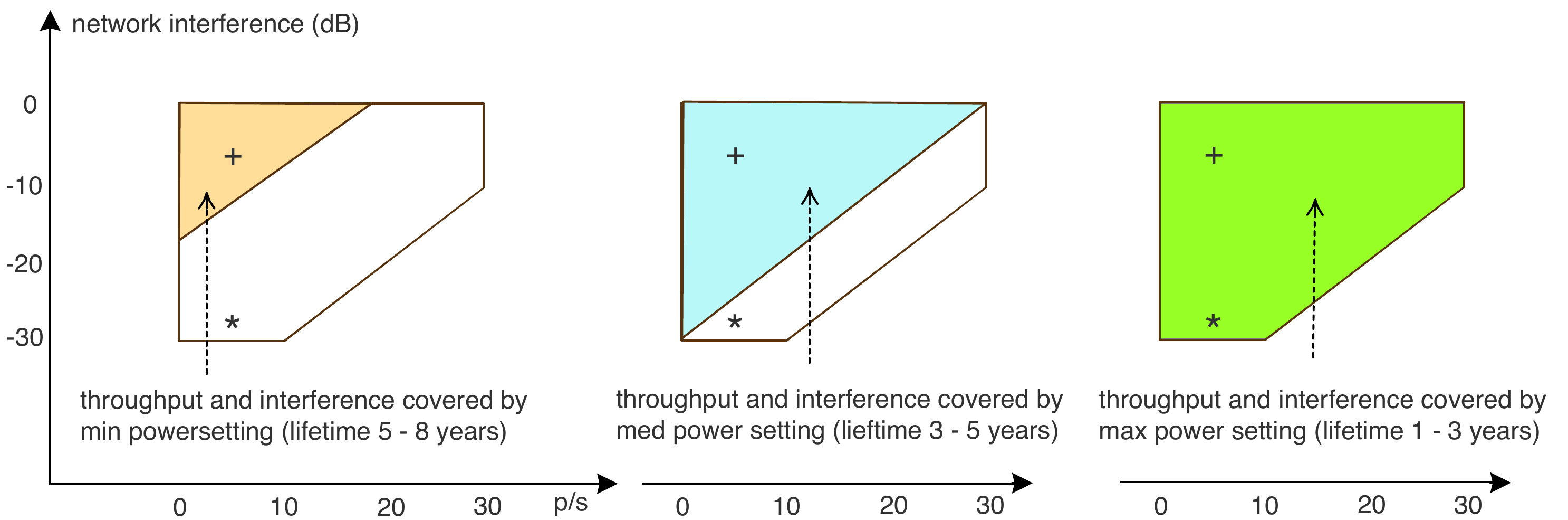}
    \caption{ODD for different configurations of the IoT system}
    \label{fig:ODD-IoT-1-2-3-separated}\vspace{-10pt}
\end{figure*}

Figure~\ref{fig:ODD-IoT-1-2-3-separated} shows the areas of the ODD covered by the three different configurations of the system (minimum, medium, and maximum power settings). Different settings can be used to cover different operational conditions. For instance, to ensure a throughput of 5 packets per second with an interference level of -28dB (working point $*$ in Figure~\ref{fig:ODD-IoT-1-2-3-separated}), the operator needs to set the power setting to maximum (setting of Figure~\ref{fig:ODD-IoT-1-2-3-separated} right). The lifetime of the network is then limited to 1 to 3 years. On the other hand, to ensure the same throughput of 5 packets per second with an interference level of -5dB (working point $+$ in Figure~\ref{fig:ODD-IoT-1-2-3-separated}), the operator can set the power setting to a minimum (setting of Figure~\ref{fig:ODD-IoT-1-2-3-separated} left). With this setting, the network's lifetime will be maximised to 5 to 8 years. 

\subsection{ODD of a Self-Adaptive System}

A human operator handles uncertainties in a traditional computing system by switching configurations. The managing system is responsible for switching configurations for different ODD regions in a self-adaptive system. 

\begin{figure}
    \centering
    \includegraphics[width = 0.9\linewidth]{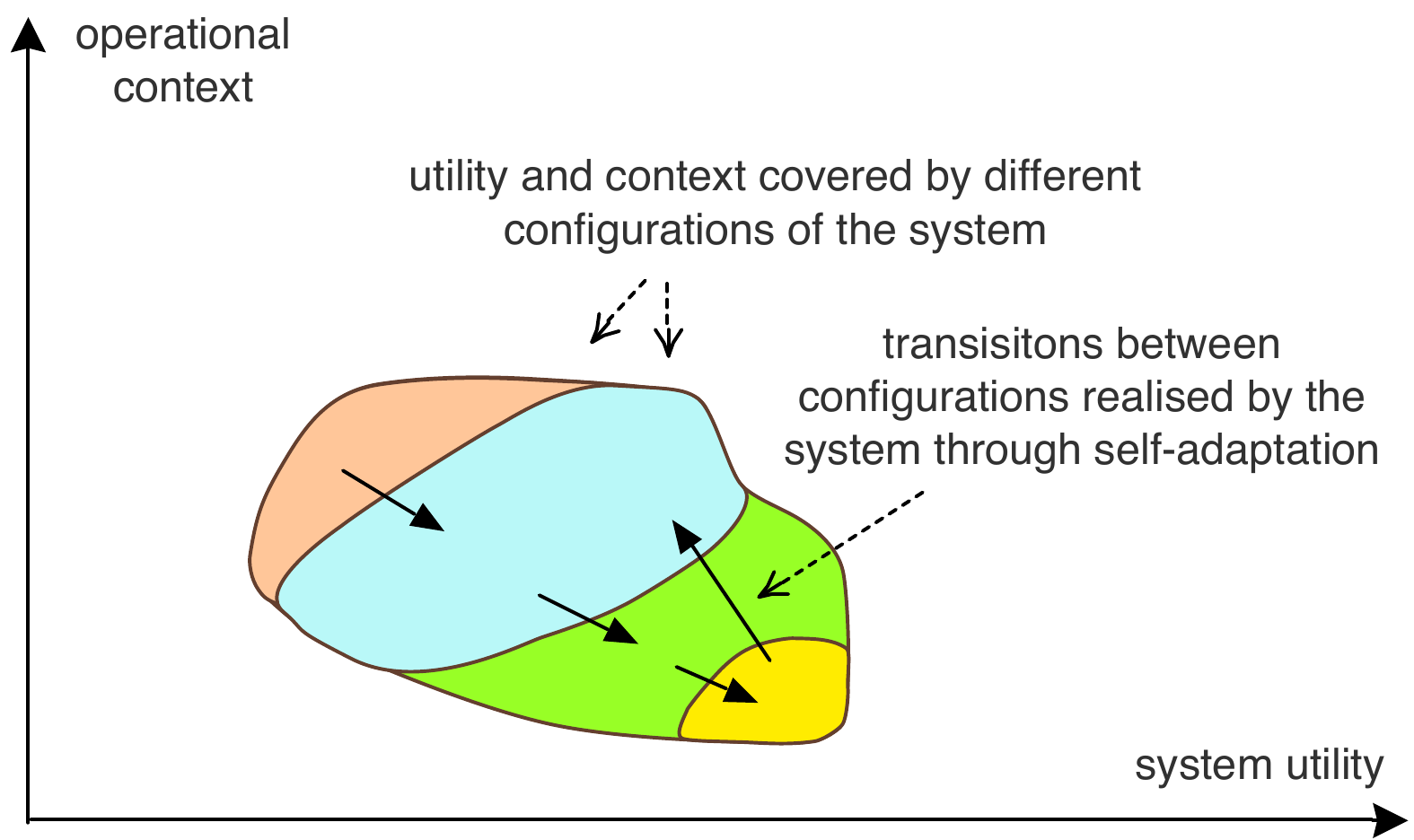}
    \caption{Switching parts of the ODD through self-adaptation}
    \label{fig:ODD-1-2-3-4}\vspace{-5pt}
\end{figure}

Figure~\ref{fig:ODD-1-2-3-4} shows how a self-adaptive system switches between the different configurations, each covering a different part of the ODD. The decision to switch from one configuration to another is the responsibility of the managing system. In general, the switch decision uses the adaptation goals that determine the system utility, the status of the managed system, the operational context in which the system operates, and possible constraints to consider.
The managing system analyses alternative re-configurations (adaptation options) as part of the decision process. For each adaptation option, the system estimates to what extent the re-configuration will achieve the adaptation goals in the given operational context. In an ODD-based approach, the managing system checks if a re-configuration switches to an ODD region with working points that satisfy the utility and operational context. The best fit of all possible re-configurations is selected (''best'' is domain-specific and determined based on stakeholder concerns).  
\vspace{5pt}\\\textbf{Example.}
Let us illustrate this for the simple IoT system. We enhance the IoT system's gateway with a managing system that: (i) tracks the managed system's status (IoT system), the current demand of throughput of the network, and the network interference, and (ii) adapts the configuration of the IoT network by switching the power setting of the motes (minimum, medium, maximum). 

\begin{figure}
    \centering
    \includegraphics[width = 0.9\linewidth]{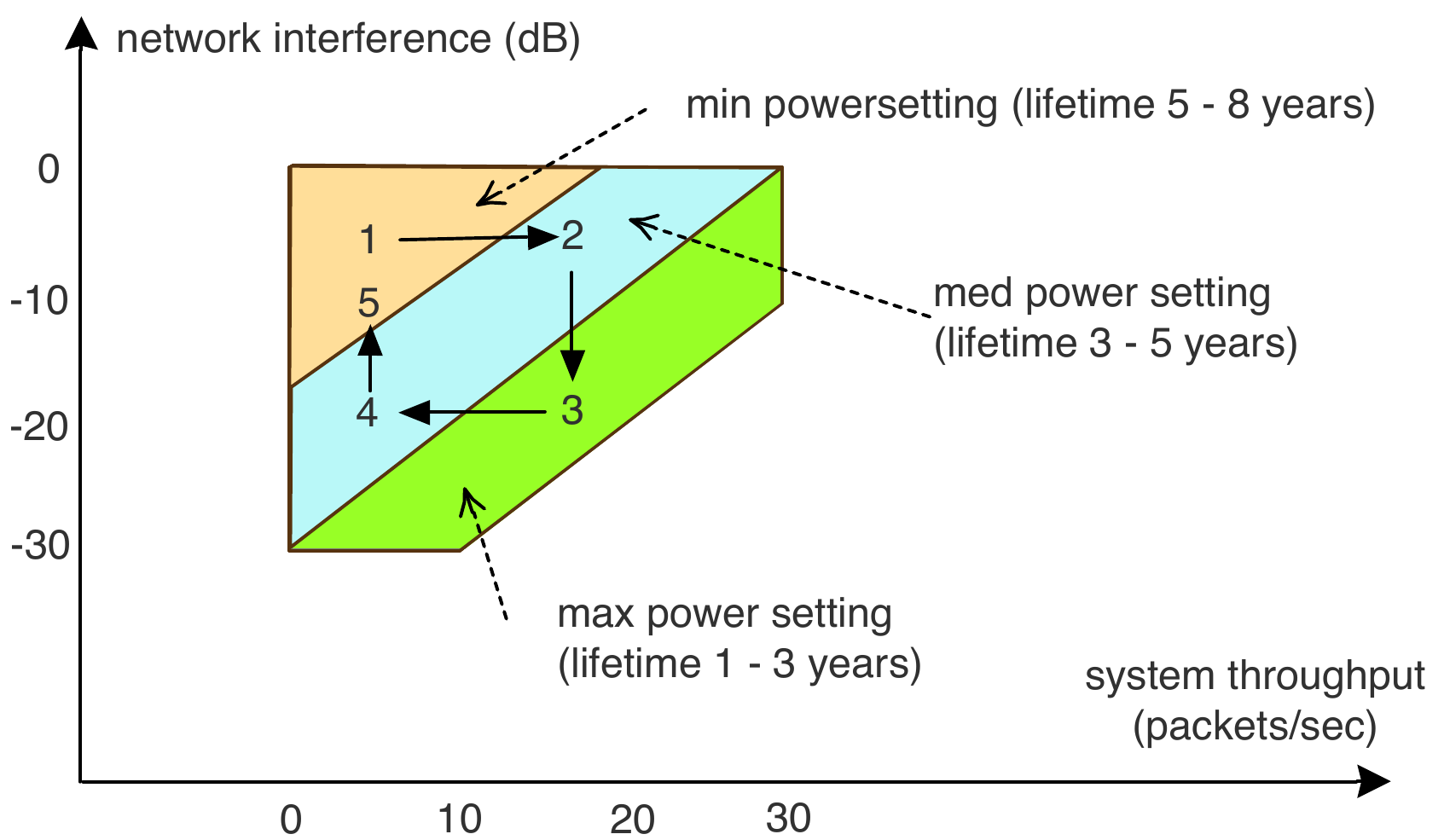}
    \caption{Self-adaptation in the IoT system}
    \label{fig:ODD-IoT-1-2-3}\vspace{-10pt}
\end{figure}

Figure~\ref{fig:ODD-IoT-1-2-3} shows how the self-adaptive IoT system switches between configurations. With a throughput demand of 5 packets/sec and a network interference of -5dB (point 1 in Figure~\ref{fig:ODD-IoT-1-2-3}), the managing system can set the power setting to a minimum, ensuring a maximum network lifetime of 5 to 8 years. When the throughput increases to 20 packets/sec (point 2), the managing system will evaluate the different options of power settings. The analysis shows that both medium and maximum power satisfy the new conditions. It will then decide to switch to the medium power setting as this will optimise the network's lifetime (3 to 5 years). When the network interference increases to -20dB (with the same demand of throughput of 15 packets/sec), the network needs to switch to the maximum power setting (point 3), reducing the network's lifetime (to 1 to 3 years). When the throughput demand decreases to 5 packets/sec (point 4), the managing system can switch back to the medium power setting, saving energy and increasing the network's lifetime. When the network interference drops to -10bB, the managing system can switch the power setting to the minimum and ensure maximum network lifetime (point 5). Hence, the self-adaptive system achieves the required throughput demand under varying network interference while maximising the network's lifetime by switching between configurations covering different ODD areas. 

\subsection{ODD and System Evolution} 

System evolution refers to an incremental development driven by changes in the environment, purpose, or use of a software system~\cite{Reussner2019}. 
Continuous evolution~\cite{RODRIGUEZ2017263} supersedes the traditional approach with periodic releases. For example, developing and releasing software using agile methods~\cite{DINGSOYR20121213} and DevOps~\cite{MISHRA2020100308} aim at increased deployment speed and higher quality. Nevertheless, software evolution remains, in essence, a human-driven process. 

\begin{figure}
    \centering
    \includegraphics[width = 0.9\linewidth]{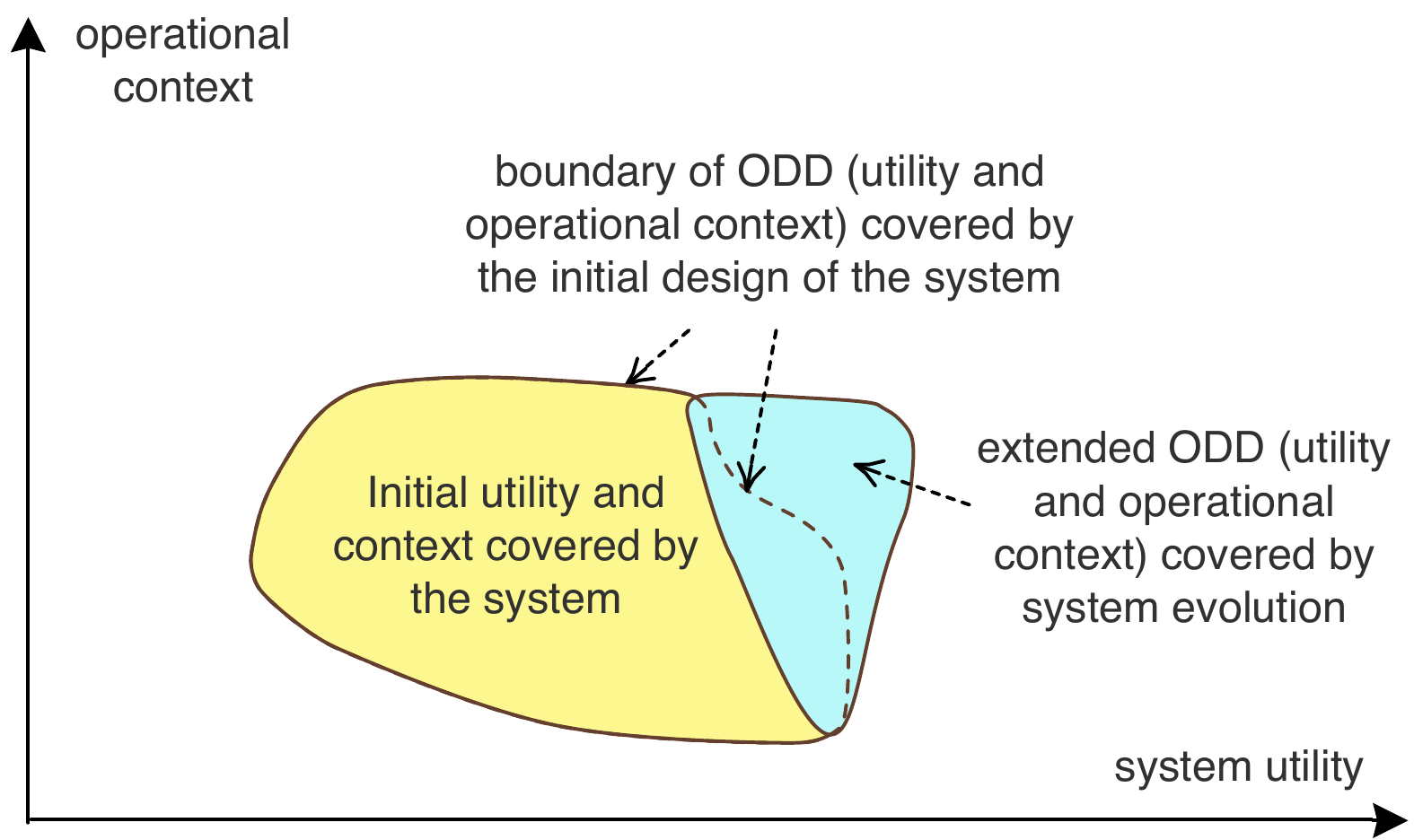}
    \caption{Extending the ODD through evolution}
    \label{fig:ODD-5}\vspace{-2pt}
\end{figure}

\begin{figure}
    \centering
    \includegraphics[width = 0.9\linewidth]{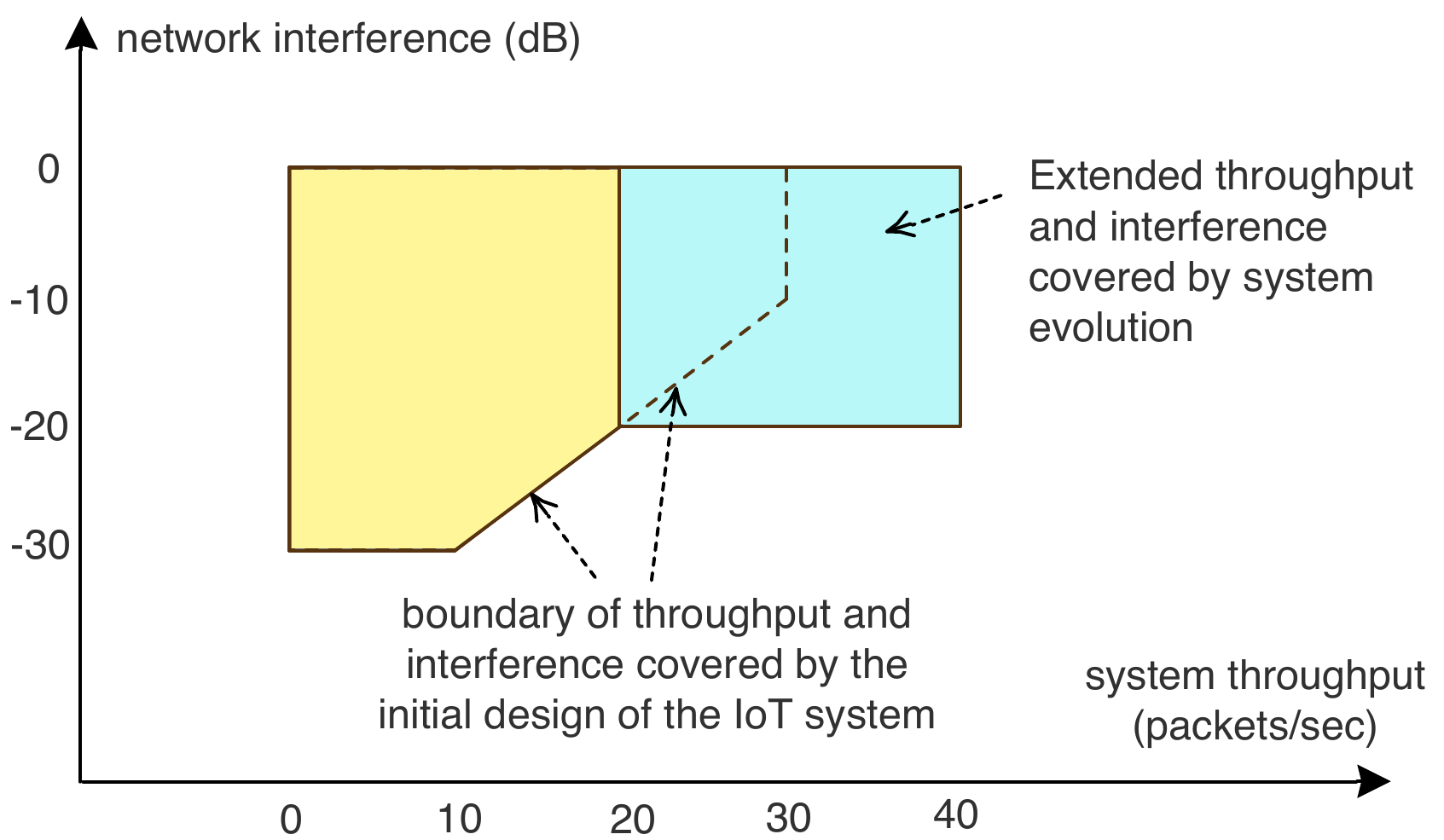}
    \caption{Extending ODD through evolution in the IoT system}
    \label{fig:ODD-IoT-4}\vspace{-5pt}
\end{figure}

Figure~\ref{fig:ODD-5} illustrates how system evolution requires an extension of the ODD.  
In this illustration, we extend the ODD with a new region that provides a system configuration with a utility and operational context the initial system's design did not cover. Such unanticipated changes require evolution. Note that system evolution may change the ODD in other ways. For instance, areas in the initial design may change, e.g., when deploying a more effective algorithm or when certain functionality is no longer required, areas may be removed.
\vspace{5pt}\\\textbf{Example.}
Figure~\ref{fig:ODD-IoT-4} illustrates how the ODD of the simple IoT system may be enhanced through evolution. 
In this example, system evolution enables the system to increase its throughput up to 40 packets per second for a network interference up to \mbox{-20dB.} Such evolution may require support for a new mechanism to adapt the IoT network, such as the ability to switch between different radio settings at the devices.  

\section{Self-Evolution} 

We now peek into the future and present a possible architecture for self-evolving systems. Central to this proposal is the role of the ODD as a representation of an evolution target that enables a system
to assess and select decision alternatives to self-evolve and deal with conditions initially not anticipated. We did not consider ODD as a driver for self-evolution in an earlier proposal~\cite{abs-2204-06825}. It is important to note that the presented architecture is explorative, even in parts, speculative. 

\subsection{ODD as Representation for Evolution Target}

We consider the case of system evolution where the systems ODD is extended to deal with new requirements and operating context. We assume that the system constraints do not change. We may develop similar reasoning for other evolution cases. Evolving a system requires knowledge about the new ODD area a system should cover, defined by a space of working points that covers a "range of utility" for a "range of contexts" not covered by the initial ODD. More precisely:\vspace{-5pt}

\begin{equation}
    ODD_{Se} = ODD_{Si} \cup ODD_{e} 
\end{equation}

That is, the operational design space $ODD_{Se}$ of the system $Se$ after evolution is equal to the union of the operational design space $ODD_{Si}$ of the system $Si$ based on its initial design and the operational design space $ODD_{e}$ of the extension by evolution. We refer to $ODD_{e}$ as the evolution target. 
\vspace{5pt}\\Self-evolution then needs to: (1) detect the need for evolution to cover $ODD_{e}$;  (2) determine a new system configuration that satisfies the evolution target $ODD_{e}$; and (3) enact the new system configuration extending $ODD_{Si}$ with $ODD_{e}$. 
%
%
The architecture for self-evolving systems supports these steps. 

\begin{figure}
    \centering
    \includegraphics[width = 0.5\textwidth]{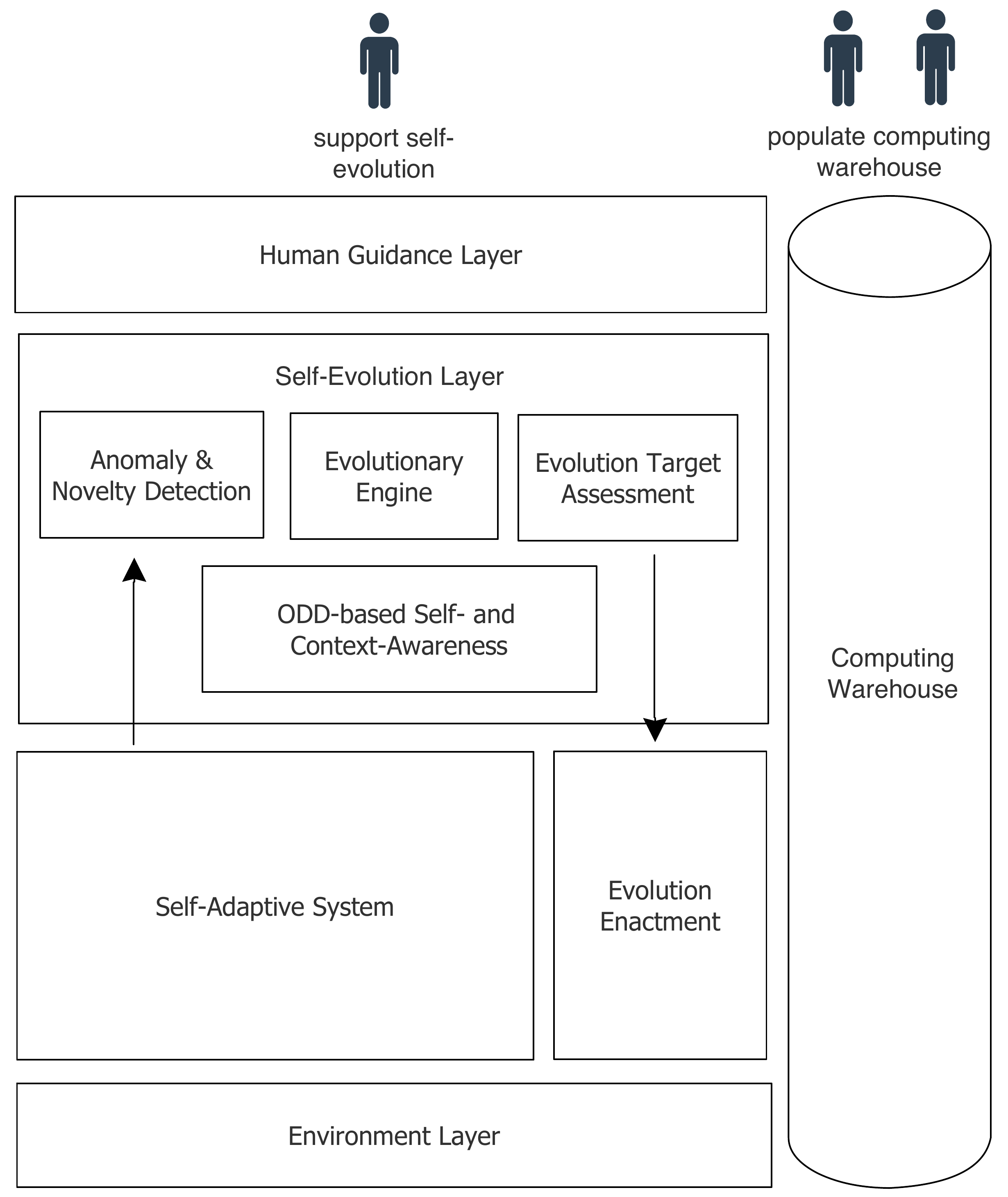}
    \caption{Conceptual architecture of a self-evolving system}
    \label{fig:self-evolution-architecture}\vspace{-10pt}
\end{figure}

\subsection{Architecture for Self-Evolution}
Figure~\ref{fig:self-evolution-architecture} shows the conceptual architecture of a self-evolving system. 
We briefly outline each architecture element's responsibilities and explain how the evolution target enables the system to decide how to evolve.
\vspace{5pt}\\\textbf{Environment layer}: the context in which the system operates that is not under the designer's control. To support self-evolution, the environment must be equipped with sensors to monitor relevant context parameters (see further below). 
\vspace{5pt}\\\textbf{Self-adaptive system}: we assume that the basic system is self-adaptive. This assumption is based on the observation that self-adaptation is nowadays widely applied~\cite{3524844.3528077,survey-report}. The self-adaptive system interacts with the environment to realise the system's requirements. The self-adaptive system can deal with anticipated changes by adapting itself when conditions change (see Figures~\ref{fig:ODD-1-2-3-4} and~\ref{fig:ODD-IoT-1-2-3} with the explanations). As such, the self-adaptive system design covers the initial ODD. To support self-evolution, the self-adaptive system needs to provide support to be monitored and evolved. Monitoring (of the self-adaptive system and the environment) should support tracking the self-adaptive system's working point within the initial ODD and detection of violations. Self-evolution enables the system to extend (change) its ODD on the fly. 
\vspace{5pt}\\\textbf{Computing warehouse}: provides computing elements that can be used by self-evolving systems to evolve during operation. We refer to these elements as \textit{auto-evolution-enabled computing elements}. Examples are software components, modules to use a service, learning algorithms, etc. To enable running systems to integrate these elements autonomously, each element is equipped with a \textit{data sheet} and a \textit{usage guide}. The data sheet provides an abstract representation of an evolution target, meaning the functionality provided by the element, a range of expected quality properties, and a range of operational contexts in which it can be used. The usage guide provides a specification for obtaining, integrating, and configuring the element. Furthermore, a computing warehouse provides a \textit{catalogue} that running systems can explore to find new elements. Computing warehouses leverage basic principles of off-the-shelf components, third-party services, and open-source software repositories. Computing warehouses can be populated with auto-evolution-enabled computing elements by software companies. They can be operated by a private company, within an ecosystem, or via an open platform. 
%
%
\vspace{5pt}\\\textbf{ODD-based self- and context-awareness}: provides first-class runtime representations of the architecture of the underlying self-adaptive system with its ODD based on the requirements of the system, constraints, and its context. Different areas of the ODD map to different configurations of the system. The system keeps track of the working point within the ODD. 
\vspace{5pt}\\\textbf{Anomaly and novelty detection}: detects unanticipated changes through anomaly or novelty detection events the system encounters or by new goals or constraints added to the system by humans (through the human guidance layer, see below). If an anomaly or novelty is detected, such unanticipated change may bring the system to a working point outside the ODD, requiring the system to go to a safe state and then take action to extend its ODD to deal with the unanticipated change. The stakeholders will specify the evolution target when adding a new system goal. When the system encounters a trigger for evolution, it will start the evolutionary engine.
\vspace{5pt}\\\textbf{Evolutionary engine and evolution target assessment}: autonomously evolves the self-adaptive system based on anomaly and novelty detection triggers. The evolution engine will evolve the architecture model of the underlying system using an evolutionary learning pipeline until the evolution target is satisfied. In this process, the engine will exploit the option to integrate new elements of computing warehouses. To that end, the engine must be equipped with a semantic resolution mechanism to enhance the current architecture with new modules by  matching the current ODD with the ODD enhanced by the evolution target. The evolution target is either provided by the stakeholders in case of a new goal and/or operating context, or it is obtained via the ODD extension of the auto-evolution-enabled element obtained from a computing warehouse to evolve the system when dealing with an anomaly or novelty trigger. Through an evolutionary learning process, the engine will evaluate the options for evolving the architecture model using a sandbox. Hereby, the engine will evaluate new candidate architectures and their corresponding ODD using the evolutionary target assessment component. The assessment will be driven by the evolution target that extends the ODD of the underlying system. The evolution engine may first explore potential re-configurations options, and once a sufficient good candidate architecture is found, further explore the evolution target (by optimising the configuration of the system model) to collect sufficient evidence that the system will behave as intended (i.e., within the given context realise the required utility under the given constraints). 
\vspace{5pt}\\\textbf{Evolution enactment}: Once a new architecture that satisfies the evolution target is identified, the new configuration is enacted. Thereby, the system may need to deploy new auto-evolution-enabled elements from a computing warehouse.  
\vspace{5pt}\\\textbf{Human guidance layer}: 
a self-evolving system may involve human experts to guide the self-evolution. A dashboard may visualise key performance indicators and enable experts to add new goals or constraints, give feedback on discovered anomalies or novelties, provide advice on architecture evolution, or support the enactment of a new system configuration. 
\vspace{5pt}\\\textbf{Example.}
In the scenario of the simple IoT system, see Figure~\ref{fig:ODD-IoT-4}, the stakeholders trigger evolution. The evolution target is defined as:
\vspace{5pt}\\
 \mbox{\ \ \ \ }$ODD_{e} = \{(u,c)~|~c \in \{0,-20\} \land u \in \{20, 40\}\}$
\vspace{5pt}\\
%
with $c$ the context in $dB$ and $u$ the throughput in packets/sec. To address this need, the system can search a computing warehouse to find a software module that supports the evolution target. An example data sheet could look like this:  
%
%
\begin{quote}
\textit{Data sheet:} $<$Frequency [(415, 868) MHz], Modulation [FM], Throughput [(0 ... 50) packets/sec], Network interference [(0 ... -30) dB] ... $>$
\end{quote}
%
Such a module would enable the system to switch between two different radio frequencies (415 and 868 MHz), while supporting the range of required throughput (0...50 packets/sec) and network interference (0...-30 dB). When the evolutionary engine identifies a semantic match of this module with the evolution target, and the module is compatible with the architecture of the devices in use (derived from the usage guide), it includes the specification of the module in the architectural model and uses a sandbox to configure the module and gather evidence that the evolution target is met for the evolved system. When this evidence is obtained, the new architecture will be enacted by collecting the module from the computing warehouse and integrating it into the running system.  


\section{Open Challenges}

\begin{figure}
    \centering
    \includegraphics[width = 0.49\textwidth]{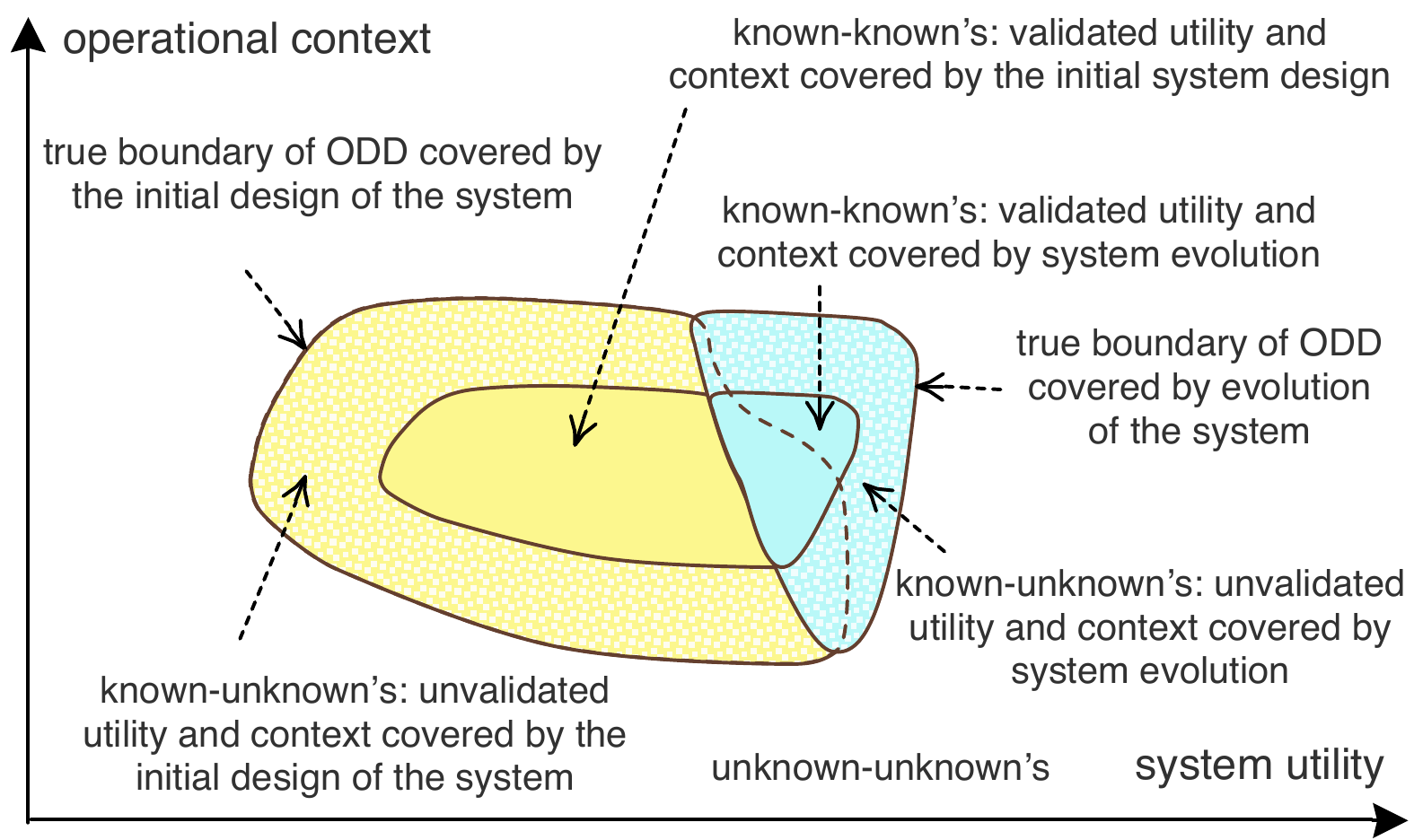}
    \caption{Degree of knowledge of regions of the OOD}
    \label{fig:ODD-5-unknowns.pdf}\vspace{-10pt}
\end{figure}

The first key challenge to realise self-evolution is establishing the concept of a computing warehouse. Central to that challenge will be the specification of the data sheet and usage guide of auto-evolution-enabled elements. Related is the challenge of specifying ODD-based self- and context awareness. In the explanation above, we assumed that the boundaries of ODD are known. Yet, in many practical systems, this is not the case. Figure~\ref{fig:ODD-5-unknowns.pdf} shows an example of the ODD of a system where some parts are known-known's, others are known-unknown's, and other are unknown-unknown's. Such ODD representations require mechanisms to express these degrees of uncertainty. Another challenge is the need for new anomaly and novelty detection mechanisms along with means to link them to the ODD and evolution target of a system. Another key challenge is a need for ontologies and semantic resolution mechanisms to support the identification of new evolving architectures. Realising suitable evolutionary engines poses another key challenge. The synchronisation of enacting a new architecture poses a significant challenge, particularly for self-adaptive systems, as it requires that the managed and managing system evolve in tandem. Lastly, mechanisms are required to support human guidance to represent and add new goals add constraints, and support the evolutionary pipeline.  

\section{Conclusions}

In this paper, we introduced the concept of ODD for a self-adaptive system. We then used the concept to explain why self-adaptation is insufficient to deal with evolution. Given the need for further automation of system evolution, we presented a conceptual architecture for a self-evolving system.  Central to this architecture is the evolution target that refers to the ODD extension of an evolving system. We explained how the evolution target enables the system to identify and decide when and how to evolve. Finally, we outlined open challenges to realise the approach for self-evolution. We believe these challenges need to be tackled jointly by researchers with complementary backgrounds and this in a step-wise manner, starting from a closed world setting, towards shared  ecosystems, and eventually an open world. 

\bibliographystyle{plain}
\bibliography{refs}

\end{document}